\documentstyle[prl,aps,epsf,rotate,multicol]{revtex}

\begin{document}
\draft

\title{Spatial Structure of the Internet Traffic}

\author{Marc Barth\'elemy}
\address{
CEA, Service de Physique de la Mati\`ere Condens\'ee\\
BP12 Bruy\`eres-Le-Ch\^atel, France}

\author{Bernard Gondran}

\address{
R\'eseau National de T\'el\'ecommunications pour\\ 
la Technologie, l'Enseignement et la Recherche\\
151, Bld de L'H\^opital, 75013 Paris, France}

\author{Eric Guichard}
\address{Equipe R\'eseaux, Savoirs \& Territoires\\
Ecole normale sup\'erieure, 75005 Paris, France}

\date{\today}
\maketitle

\begin{abstract}
The Internet infrastructure is not virtual: its distribution is
dictated by social, geographical, economical, or political
constraints. However, the infrastructure's design does not determine
entirely the information traffic and different sources of complexity
such as the intrinsic heterogeneity of the network or human practices
have to be taken into account. In order to manage the Internet
expansion, plan new connections or optimize the existing ones, it is
thus critical to understand correlations between emergent global
statistical patterns of Internet activity and human factors. We
analyze data from the French national `Renater' network which has
about $2$ millions users and which consists in about $30$
interconnected routers located in different regions of France and we
report the following results. The Internet flow is strongly localized:
most of the traffic takes place on a `spanning' network connecting a
small number of routers which can be classified either as `active
centers' looking for information or `databases' providing
information. We also show that the Internet activity of a region
increases with the number of published papers by laboratories of that
region, demonstrating the positive impact of the Web on scientific
activity and illustrating quantitatively the adage `the more you read,
the more you write'.
\end{abstract}


\begin{multicols}{2}


\section{Introduction}

Internet connects different routers and servers using different
operating systems and transport protocols.  This intrinsic
heterogeneity of the network added to the unpredictability of human
practices \cite{Huberman97} make the Internet inherently unreliable
and its traffic complex
\cite{Leland94,Csabai94,Thompson97,Feldmann98,Takayasu00}.

There has been recently major advances in our understanding of the
generic aspects of the Internet
\cite{IMP,Faloutsos99,Caldarelli00,Pastor01} and web
\cite{Kumar99,Broder00,Albert99,Adamic99,Huberman98,Mossa02} structure
and development. Concerning data transport, most of the studies focus
on properties at short time scales or at the level of individual
connections \cite{Leland94,Crovella97,Willinger97}, while studies on
statistical flow properties at a large scale
\cite{Csabai94,Thompson97,Takayasu00,Fukuda99} concentrate essentially
on the phase transition to a congested regime. Despite of these
results, large scale studies of traffic variations in time and space
are still needed before understanding the new social practices of
Internet users.

In this paper, we study the spatial structure of the large scale flow. We
present in part II the data studied and in parts III and IV the
results of our analysis, showing the existence of a spanning network
concentrating the major part of the traffic. Finally, in part V we
relate the flow properties and its spatial distribution to scientific
activity measured by the number of published papers.


\section{Data studied}

An important difficulty is to obtain real data measurements of the
Internet traffic on a global scale. The availability of data of the
French network `Renater' allows us to consider the cartography of
Internet's traffic and its relation with regional socio-economical
factors.

\begin{figure}
\narrowtext
\vspace*{1cm}
\centerline{
\epsfysize=0.7\columnwidth{{\epsfbox{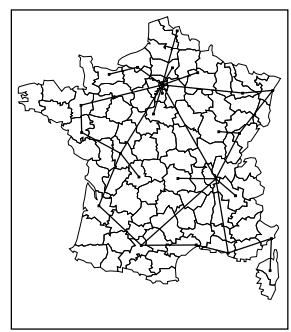}}}}
\vspace*{.5cm}
\caption{ The Renater Network. There is a total of about $30$
interconnected main routers. To each router corresponds a region
comprising a main city. The data consist in a flow matrix $F_{ij}(t)$
which gives the effective flows (virtual or physical) on the connection
between routers $i$ and $j$. For more details on this network, see the
web page {\sf http://www.renater.fr} and for an animated version of
flows, see {\sf http://barthes.ens.fr/metrologie/Renater01}.}
\label{Renater}
\end{figure}

The French network `Renater' has about $2$ million users and is
constituted of a nation-wide infrastructure and of international links
\cite{Renater}. Most of the research, technological, educational or
cultural institutions are connected to Renater
(Fig.~\ref{Renater}). This network enables them to communicate between
each other, to get access to public or private world-wide research
institutes and to be connected to the global Internet.

We first restrict our analysis to the national traffic and exclude the
information exchange with external hosts and routers such as US and
Europe Internet or peering with other ISPs. This restriction to a
small part of the Renater traffic ($\sim 5\%$ of roughly $2000$
Gigabytes a day) has two methodological advantages: First, it ensures
that the traffic studied is strictly professional (mail to
non-academics, like family, friends, consultation of newspapers,
e-commerce, etc. goes through outer ISP and is not taken into
account); Second, it helps to understand the regional traffic
structure and its relation with local economical factors. We believe
that the global patterns emerging for the Renater network will be
relevant for larger structures such as the global Internet.

The data consist of the real exchange flow (sum of Ftp, Telnet, Mail,
Web browsing, etc.) between all routers, even if there is not a direct
(physical) link between all of them. For a connection $(i,j)$ between
routers $i$ and $j$ ($i\neq j$), $F_{ij}(t)$ (in bytes per $5$
minutes) is the effective information flow at time $t$ going out from
$i$ to $j$. For technical reasons, data for a few routers were not
reliable and we analyzed data for $26$ routers which amounts in
$26\times 25$ matrices $F_{ij}(t)$ given for every $\Delta t=5$
minutes for a two weeks period (the quantities $F_{ii}$ are excluded
from the present study).

\begin{figure}
\narrowtext
\centerline{
\epsfysize=0.6\columnwidth{{\epsfbox{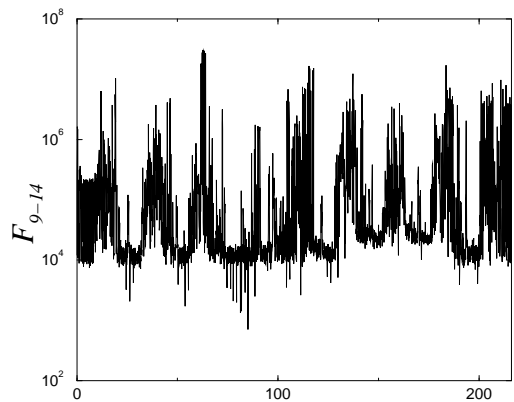}}}
}
\vspace*{.5cm}
\caption{ Lin-Log plot of the information flow versus time from
Grenoble to Marseille over a $9$ days period. We represent the raw
data: the number of bytes (per $5$ minutes) exchanged between these
two cities.  One can see the different days and within days, bursts of
intense activity.}
\label{flux}
\end{figure}

As an example of the measured time-series, we show (Fig.~\ref{flux})
the information flow versus time between two routers located in
Grenoble and Marseille for a nine days period. One can see the
different days and within days, bursts of intense activity. In this
study, we focus on the flow and not on the growth rate (and its
correlations) used in a previous study \cite{Barthelemy02}.


\section{Databases versus Active Centers}

We now present our empirical results. The time averaged incoming flow
\begin{equation}
F_{in}(i)=\sum_j\overline{F}_{ji}
\end{equation}
at a given router $i$ is a measure of the Internet activity of the
corresponding region (the over-bar denotes the average over time). On
the other hand, the average outgoing flow
\begin{equation}
F_{out}(i)=\sum_j\overline{F}_{ij}
\end{equation}
can be interpreted as the total request emanating from other
routers. It is thus a measure of the degree of interest produced by
this region.


We plotted both quantities $F_{in}$, $F_{out}$ versus their rank
(Fig.~3a,b). In contrast with many cases observed since the work of
Zipf \cite{Zipf49}, the observed distributions are not power laws but
exponentials. This might be the signature of a transient regime and
would mean that the Internet didn't reach his stationary state, but it
is more probably the sign that the Internet traffic has a unique, non
hierarchical-type structure \cite{Marsili98}. This exponential
behavior also means that\---at least in the Renater network\---there
are essentially two categories of regions. Considering Internet
activity (Fig.~3a), one can distinguish active from (almost) inactive
regions. Roughly, there are about eight cities which receives $80\%$
of the total traffic, the rest being (exponentially)
negligible. Concerning the outgoing flow (Fig.~3b), there are about
five most visited regions, the rest being comparatively
`unattractive'. We checked that for different time windows the order
of these cities can slightly change, but the exponential behavior is
independent of these seasonal effects.

It is interesting to note that the most active and visited regions are
not the same showing that each region has its specific
activity. Regions with a large incoming flow can be classified as
active research centers with a great need of information, and regions
with a large outgoing flow correspond to important information
resources such as e.g., databases or libraries.


At this stage, we have shown that in the Renater traffic there is a
small number of receivers (located in active regions) and emitters
(visited databases). However, a further question concerns the
secondary routers and the fine structure of flows. Indeed, $F_{in}$
(and similarly for $F_{out}$) could be a sum of many small
contributions coming from many regions, or in contrast there could be
only few regions which exchange a significant flow. Simple quantities
which can characterize the fine structure of the incoming flow at
router $i$ are the $Y_k$'s introduced in another context
\cite{Derrida}
\begin{equation}
Y_k(i)=\sum_{j=1}^N (W_{ji})^k
\end{equation}
where $W_{ji}=F_{ji}/F_{in}$ is the weight associated with the
incoming flow $F_{ji}$ (and similar expressions for the structure of
outgoing flow). It is easy to see that $Y_0=N$, $Y_1=1$ and the first
non trivial quantity is $Y_2$. We can illustrate the physical meaning
of $Y_2$ with simple examples. If all weights are of the same order
$W_{ji}\sim 1/N$ for all $i,j$ then $Y_2\sim 1/N$ is very small. In
contrast, if one weight is important for example of the order $\sim
1/2$ and the others negligible $\sim 1/2(N-1)$ then $Y_2\sim 1/4$ is
of order unity. Thus $1/Y_2$ is a measure of the number of important
weights. We plot $Y_2$ for both the incoming and outgoing flows (the
statistics is over two weeks). The result (Fig.~\ref{y2}) shows
clearly that the most probable value is $1/4$ and that $Y_2$ is larger
than $1/N\simeq 1/30\simeq 0.03$ (except for few cases which appear in
the histogram).  This confirms the fact that a few routers are
exchanging most of the information, the rest of the network being
negligible.


\section{Spanning network}

In order to illustrate the above results, we construct the network
$S_k$ connecting a number $k$ of routers and carrying the maximal flow
denoted by $F(S_k)$. We increase $k$ from $2$ to $26$ and we obtain
the result plotted in Fig.~(\ref{spanning}a). It appears clearly that
a small fraction of links carry most of the flow. This behavior is
encoded in the fact that $F_{tot}$ is a power law for $p\to 0$
\begin{equation}
F_{tot}(S_k)\sim p^{\theta}
\end{equation}
with an exponent smaller than one ($\theta\simeq 0.6$), so that a
small variation of the number of connections leads to a large
variation of the transported flow.

This analysis completes the Renater traffic map: there is a small
number of receivers and emitters exchanging significant information
between them, the rest of the network being exponentially
negligible. This demonstrates the existence of a `spanning network'
carrying most of the traffic and connecting the main emitters to the
main receivers.

In order to visualize this spanning network on the French map, we
filter the flow with the following procedure. We first select flows
above a certain threshold $F_c$, and then we select a connection
$(ij)$ only if the corresponding flow $F_{ij}$ represents a large
percentage of (i) the outgoing flow from $i$ and (ii) the incoming
flow in $j$. The result is shown on Fig.~\ref{spanning}b. We checked
that the instantaneous (or averaged over a different time window)
spanning network is eventually different, but always with the same
characteristics (small number of interconnected emitters and
receivers). The procedure described above could thus be used as a
simple filter in order to visualize in real-time a complex flow
matrices.


\section{Digression to Scientometrics: Internet Traffic 
versus Scientific activity}

So far, we have studied statistical properties of the traffic, but an
important point is to relate them to economical or social factors. The
Internet activity should in principle be related to social pointers
such as the number of inhabitants, the number of students, and so
on. From our data, the indicator which shows the best correlation with
Internet activity is scientific activity measured by the number of
published papers \cite{Braun93}. One can expect that the more a
scientist consults books or data, the more he/she will publish. This
principle, `the more you read, the more you write', although commonly
accepted in a number of historical cases \cite{Baratin96}, is
difficult to evaluate quantitatively. The main difficulty being the
measure of the amount of information gathered by scientists in
libraries. In the case of Internet, the information gathered by
scientists working in a given region can be estimated by the average
incoming flow in the corresponding router. Since the information
needed for a scientist is usually scattered world-wide, it is
important here to take into account the total incoming flow, including
exchanges with international hosts. We thus compare the total average
incoming flow (per scientist) with the average number of papers
published (per scientist) per year by the region's universities
(obtained from the SCI database). As a representative panel, we choose
to use data about papers published only by scientists in the national
research institution (CNRS \cite{cnrs}). We represent these data on
Fig.~\ref{flux.papers}. This plot shows that the average incoming flow
per scientist $\overline{f}$ in a region is increasing with the number
of published scientific papers per scientist $p$ by this region's
laboratories roughly as a power law
\begin{equation}
{\overline f}\sim p^\beta
\label{read}
\end{equation}
with exponent $\beta\simeq 1.1\pm 0.1$. This result confirms
quantitatively the intuitive principle stated above and is
particularly interesting from the point of view of the Web's social
impact. Indeed, it implies that the number of publications is growing
with the incoming flow as a power law with exponent $1/\beta \simeq
1$: the more one uses Internet, the more one publishes! This result
indicates that on average the use of Internet has a positive impact on
research productivity.


\section{Conclusion}

In summary, we have shown that the major part of the traffic takes
place only between a few routers while the rest of the network is
almost negligible. We have proposed a simple procedure to extract this
(bipartite) spanning network, which could have some implications in
visualization and monitoring of real-time traffic. In addition,
resources allocation and capacity planning tasks could benefit from
the knowledge of such a spanning network. These results point towards
new ways of understanding and describing real-world traffic. In
particular, any microscopic model should recover these statistical
properties and our results provide a quantitative basis for modeling
the dynamics of information flow.


We also have shown that the scientific activity of a region is
increasing with its Internet activity. This indicates that it is
difficult for a scientist to avoid the use of Internet without
affecting his/her productivity measured in terms of publications. This
result also demonstrates that in addition to increase people's social
capital \cite{Wellman01} the Internet has a measurable positive impact
on research production. More generally, it underlines the importance
of Internet as knowledge sharing vector. This study also suggests that
the Internet activity could be used as an interesting new
socio-economical pointer well adapted to the information society.


Finally, these results exhibit some global statistical patterns
shedding light on the relations between the Internet and economical
factors. It shows that in addition to the structural complexity of the
web and the Internet, the traffic has its own complexity with its own
cartography.



\begin{figure}
\narrowtext
\vspace*{0.5cm}
\centerline{
\epsfysize=0.5\columnwidth{{\epsfbox{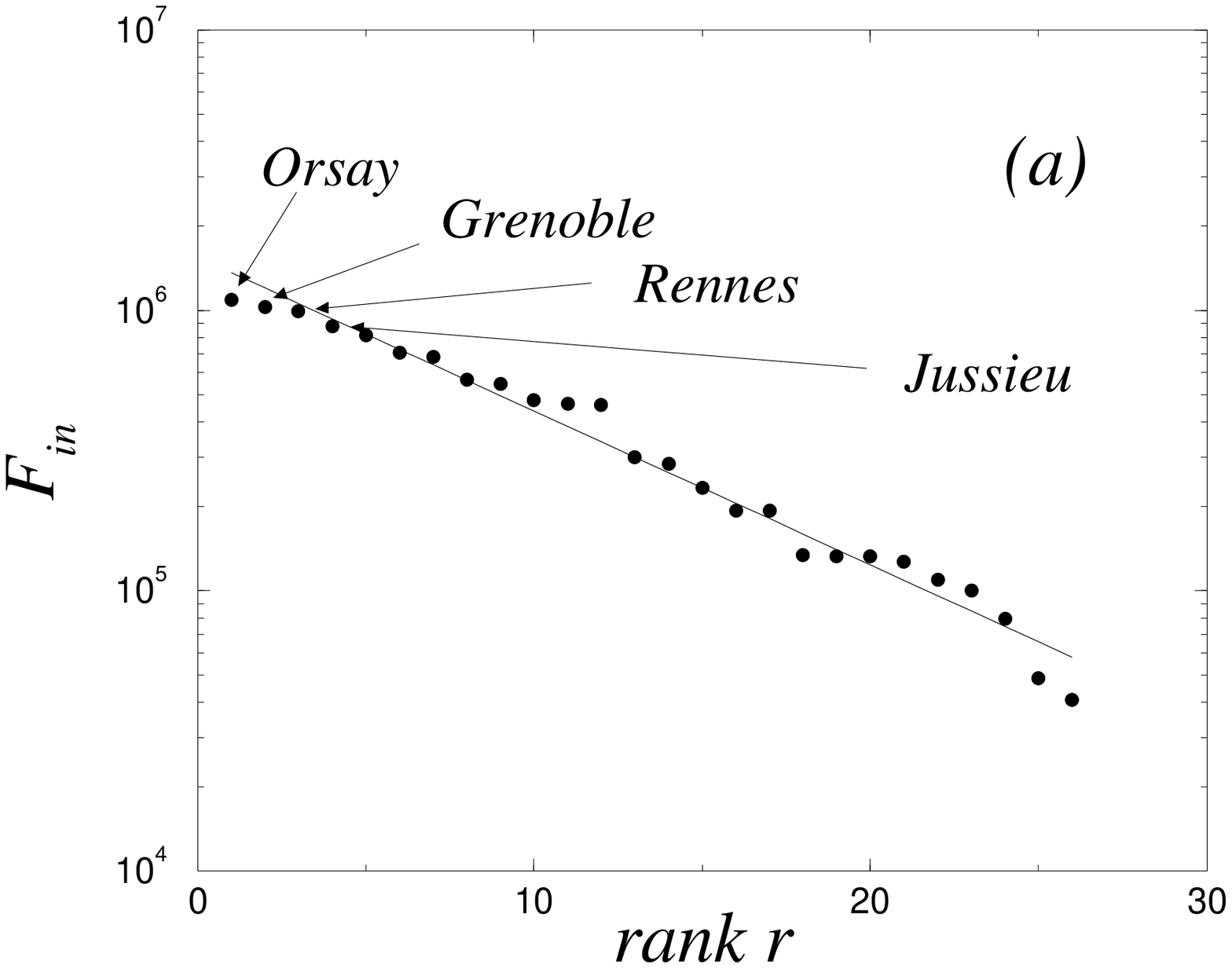}}}
\hspace*{.5cm}
\epsfysize=0.5\columnwidth{{\epsfbox{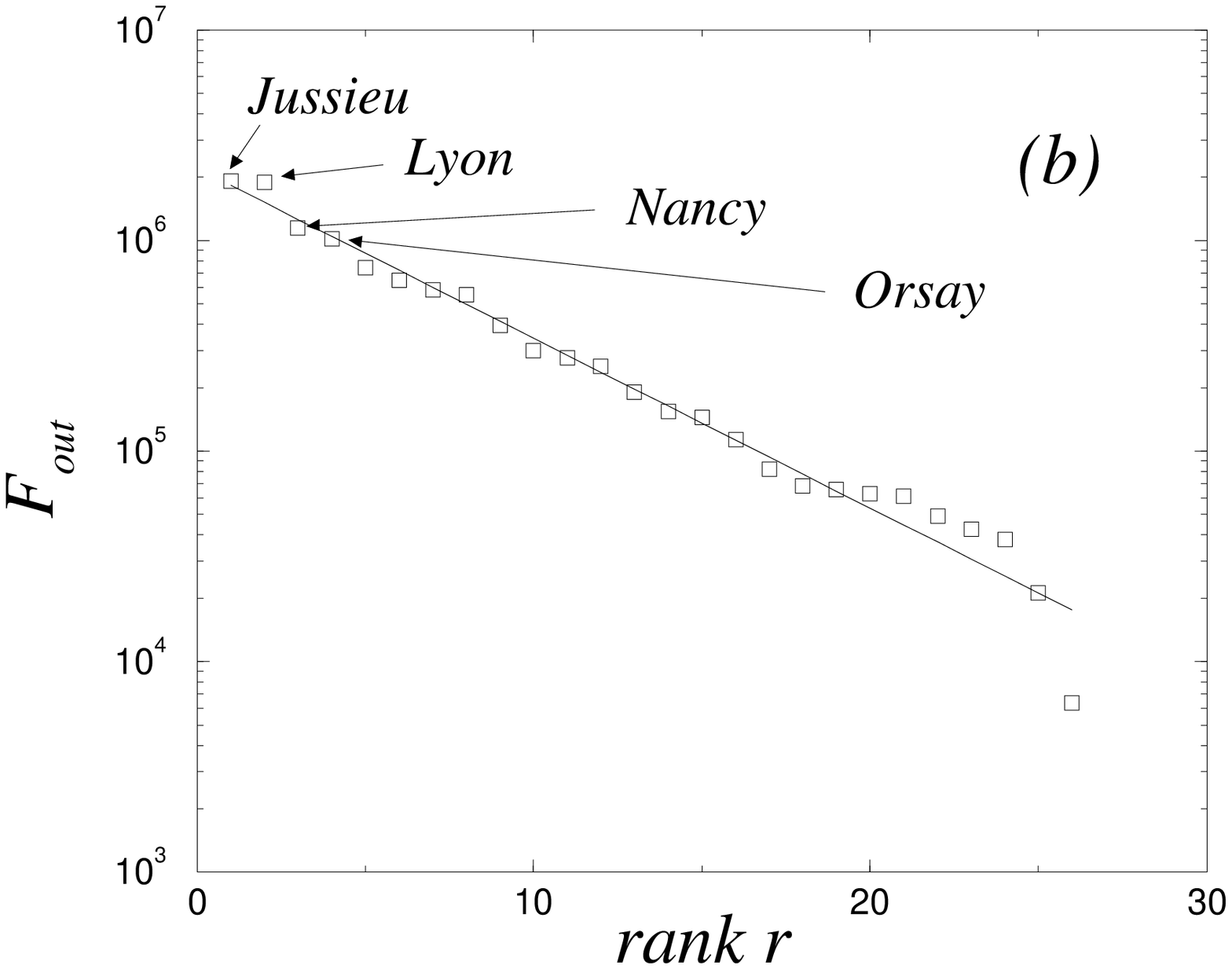}}}}
\vspace*{.5cm}
\caption{ Exponential disparities between regions. Lin-Log Zipf (rank)
plot of the Internet activity (in bytes$/5$minutes) {\bf a} Incoming
flow $F_{in}$ and {\bf b} outgoing flow, averaged over two weeks. The
relation is an exponential (indicated by solid lines)
$F_{in(out)}=exp(-r/r_{in(out)})$ with $r_{in}\simeq 8$ in the case
({\bf a}) and $r_{out}\simeq 5$ for case ({\bf b}). This result allows
one to separate regions in two distinct groups: `active'
($r<r_{in(out)}$) and `inactive' ones ($r>r_{in(out)}$). We indicate
in each case, the four first cities. This order can change according
to different period, but the exponential behavior will still holds.}
\label{fluxave}
\end{figure}


\begin{figure}
\narrowtext
\vspace*{0.5cm}
\centerline{
\epsfysize=0.6\columnwidth{{\epsfbox{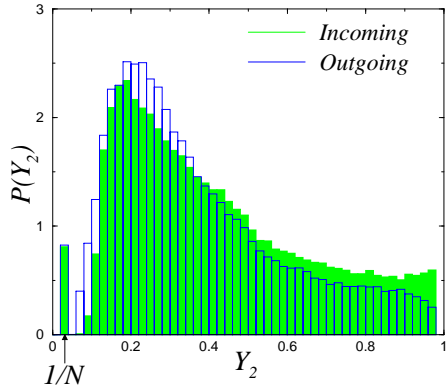}}}
}
\vspace*{.5cm}
\caption{ Fine structure of flows. We represent the probability
distribution of $Y_2$ for both incoming and outgoing flows. A small
value of $Y_2$ corresponds to a very `fragmented' flow, while a large
value means that there are only one or two important contributions,
the rest being negligible. The arrow indicates the value $1/N\simeq
0.03$ which corresponds to flows for which each router
contributes. The distributions are peaked for $Y_2\simeq 0.2$ and are
concentrated in the range $0.2-1$. This indicates that essentially a
few routers contribute to each flow.}
\label{y2}
\end{figure}


\begin{figure}
\narrowtext
\vspace*{1cm}
\centerline{
\epsfysize=0.5\columnwidth{{\epsfbox{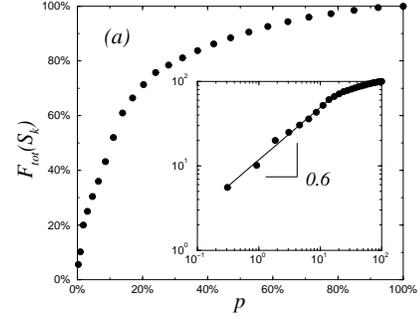}}}
}
\centerline{
\epsfysize=0.6\columnwidth{{\epsfbox{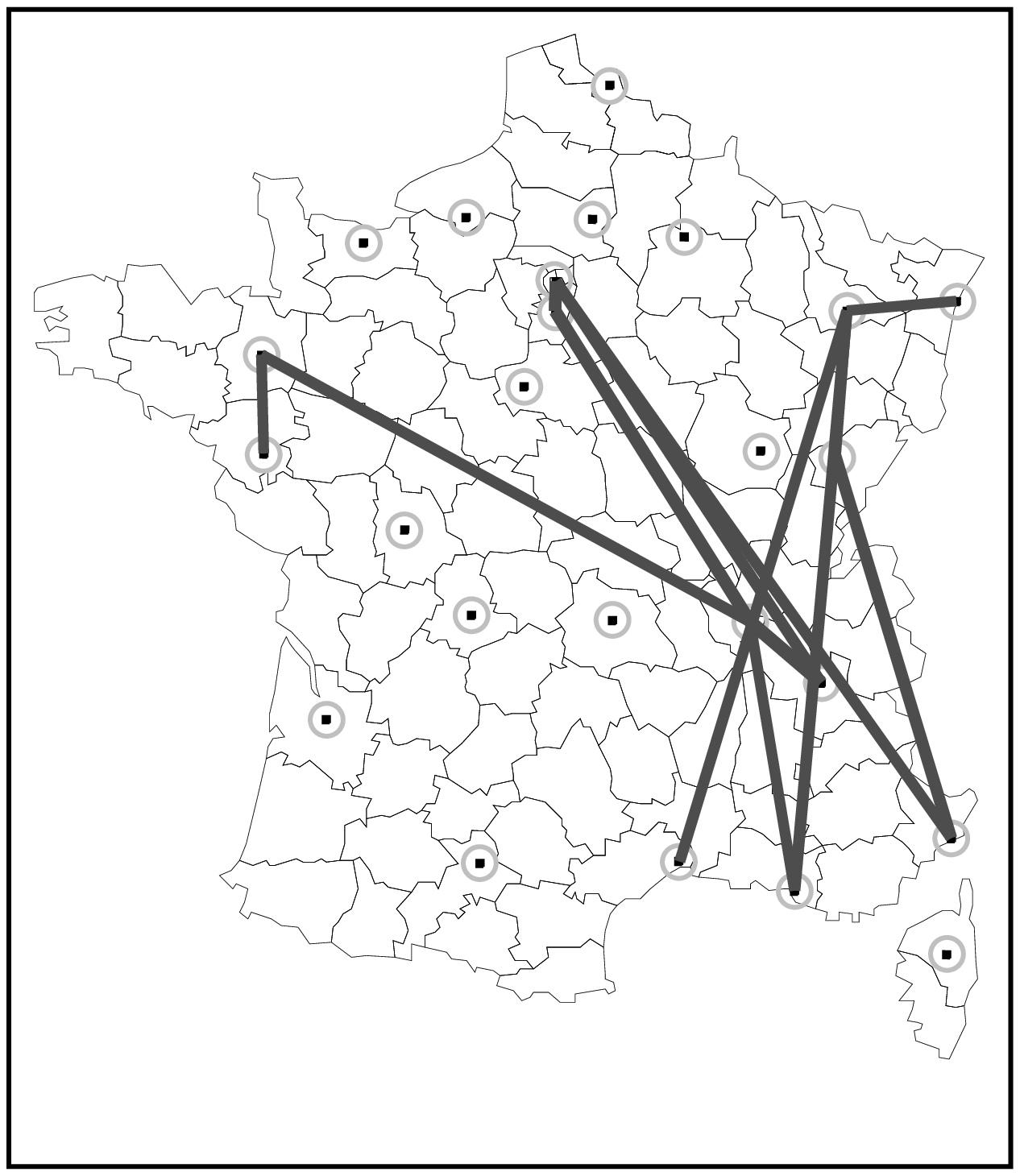}}}
}
\vspace*{.5cm}
\caption{ The spanning network. {\bf a} $F_{tot}(S_k)$ is the total
flow carried over the network $S_k$ which (i) connects a number $k$ of
routers and (ii) carries the maximal flow. The quantity $p$ is the
corresponding number of connections between $k$ routers over the total
number of possible connection in the whole network. In insert, we show
the same plot in Log-Log showing that for $p\simeq 0$, the flow is
growing as a power law $p^{\theta}$ with $\theta\simeq 0.6$. {\bf b}
We apply the filtering procedure explained in the text and we obtain
the spanning network. In this example, it is constituted by $14$
connections between $11$ routers (which is $2\%$ of the total number
possible connections) and carries $30\%$ of the total flow. On this
map, the width of the connection is a slowly increasing function of
the volume flow passing in it.}
\label{spanning}
\end{figure}


\begin{figure}
\narrowtext
\vspace*{1cm}
\centerline{
\epsfysize=0.6\columnwidth{{\epsfbox{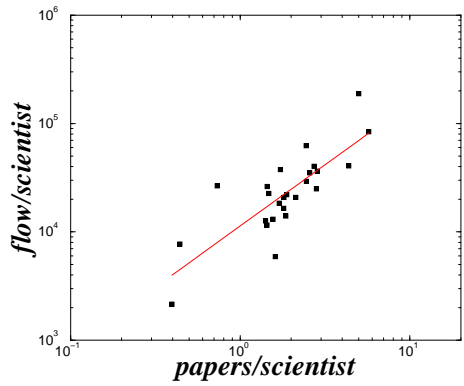}}}
}
\vspace*{.5cm}
\caption{ Internet activity versus scientific research. Internet
activity versus scientific research. Log-Log plot of the incoming flow
per scientist in a region versus the average ratio of scientific
papers per scientist published (per year) by scientists working in
that region. The solid line is a least square estimate with a slope of
order $1.1\pm 0.1$. The correlation coefficient $C=0.80$ while $F=41$
for one degree of freedom. This plot shows that scientific
productivity is increasing with Internet activity.}
\label{flux.papers}
\end{figure}


\end{multicols}

\end{document}